\documentclass[twocolumn]{jpsj2}
\usepackage{graphicx}

\def\runtitle{Morphology of Graphitic Carbon}
\def\runauthor{Masahiko \textsc{Hayashi}}

\title{%
Topological Defects and Morphology of 
Graphitic Carbon Materials: \\
An Approach Based on Differential Geometry
}

\author{%
Masahiko \textsc{Hayashi}\thanks{hayashi@cmt.is.tohoku.ac.jp}
}

\inst{%
Graduate School of Information Sciences, Tohoku University\\
Aramaki, Aoba-ku, Sendai 980-8579, Japan}

\recdate{\today}

\abst{%
It has been known that pentagons and heptagons in 
hexagonal graphitic network give rise 
to a certain amount of curvature in the 
three dimensional structure of 
graphitic carbon materials. 
The amount of curvature is quantized due to the 
symmetry of graphite and, as a result, 
the structure formed by the network is also 
restricted. 
We clarify the effects of curvature quantization 
on the forms of graphitic carbon materials, 
employing the knowledge of differential geometry, 
especially the Gauss-Bonnet theorem. 
}

\kword{%
nanotube, fullerene, graphite, topological crystal, disclination, 
differential geometry
}

\begin{document}
\sloppy
\maketitle

\section{Introduction}
After the discovery of Fullerenes 
\cite{Kroto,Osawa,Iijima_C60,Curl-Smally,Kroto_Review} 
and nanotubes \cite{Iijima_nanotube},
graphitic carbon materials have been studied intensively. 
Until now many intriguing structures have been discovered: 
larger Fullerenes \cite{Kroto_Cn,Kratschmer}, nanohorns 
\cite{Iijima_curvature}, 
kinked nanotubes \cite{Iijima_single,Yao}, Y-junctions 
\cite{Yao,Satishkumar} and so on. 
The discovery of these materials has been stimulated 
interests in more exotic structures, 
such as negatively curved periodic minimal surface 
\cite{Kroto-McKay,McKay-Terrones,Lenosky}. 

From the very beginning of the research of these materials, 
the geometrical methods have been 
powerful tools to predict, investigate and clarify 
the new structures of graphitic carbon materials. 
For example, based on the Euler's formula, 
Iijima was able to 
guess one of the basic characters of 
Fullerene, {\it i.e.}, the existence of 
twelve pentagons (five-membered rings), 
before the clarification of the C$_{60}$ 
structure \cite{Iijima_C60}. 

In this paper, aiming to apply the method 
of differential geometry to the analysis of 
more complicated topologically nontrivial 
structures, 
we develop a simple geometrical 
model for the graphitic carbon materials. 
Basic point of our model is the assumption 
that the graphite sheet is 
free from expansion and contraction; 
it can only be bent without changing 
the area of any part of the sheet. 
In order to understand this, one can 
imagine a sheet of ordinary paper, 
where expansion or contraction is much 
more difficult than bending. 
Mathematically, this kind of surface is 
called {\it developable surface}; 
when a part of the surface is 
cut out, it can be 
developable on a plane without 
expansion or contraction. 
For example, the side of a cylinder or 
a cone is a developable surface, 
whereas a part of a sphere is not. 
Whether a surface is really 
\lq\lq curved\rq\rq  like spheres or 
is just a bent plane like sides 
of cylinders or cones 
is distinguished by introducing the so-called 
{\it Gaussian curvature} 
in the framework of the differential 
geometry 
(see text for details). 
In the case of graphitic carbon materials, 
non-zero Gaussian curvature arises only 
from the {\it disclinations}, 
which correspond to the pentagons and 
heptagons in the graphitic network.  
The Gaussian curvature caused by the 
disclinations is quantized due to the 
hexagonal symmetry of the graphite sheet 
and the possible forms of the 
graphitic carbon materials are restricted by 
this effect. 
The main part of this paper is therefore 
devoted to the investigation of 
the relation between the quantized 
Gaussian curvature and the structures of the 
graphitic carbon materials 
for several simple cases, 
such as cones, spheres, kinked tubes 
and so on. 

Here we would like to note that today the 
crystals with topologically 
interesting forms are not limited to 
the graphitic carbon materials. 
Recently, \lq\lq nanotubes\rq\rq are 
found in several other materials, 
such as MoS$_2$ \cite{MoS2}, MgB$_2$
and NbSe$_2$\cite{Tanda_private}.  
Rings and M\"obius strip-type crystals 
are also found for 
NbSe$_3$\cite{Tanda_NbSe3}.
These facts suggest that 
\lq\lq topological crystals\rq\rq are 
commoner than we expect. 
We believe that the present approach will also shed 
a new light on these new materials 
in the future. 

This paper is organized as follows: 
in Sec.\ref{disclination}, we describe the basic 
properties of the disclinations in graphite 
sheet and show how the Gaussian curvature 
per one disclination is quantized. 
In Sec.\ref{G-B_application}, we apply the 
Gauss-Bonnet theorem to the cases of Fullerenes, 
Cones, kinked tubes and so on, and 
clarify how the quantized Gaussian curvature 
affects the structure of the graphitic carbon 
materials. 
In Sec.\ref{Discussion}, we give some discussion 
on the experimental observation of the 
prediction of this paper. 
The possible development of 
our treatment is also discussed. 
In Sec.\ref{Summary}, 
we summarize our results. 
In Appendices, we introduce the basic 
framework of the differential geometry, 
which is employed in this paper. 

\section{Disclination in Graphitic Network}
\label{disclination}

First we introduce the basic assumption of our model. 
Generally, the graphite sheet is hard against 
expansion and contraction, 
since the $sp^2$ bonding of 
carbon is strong and the 
bond length is almost fixed.  
The shear deformation is also 
suppressed due to the 
firmly fixed bond angle. 
Therefore the deformation arrowed 
for a graphite sheet is only 
the bending without changing 
the area of any part 
of the graphitic network. 
This means, from the knowledge of 
the differential geometry, 
that the graphite sheet forms a 
developable surface 
or, equivalently, the Gaussian 
curvature $K$ of the graphite sheet 
vanishes (see Appendix for more details). 
If we employ a more easy analogy, 
the graphite sheet is like a sheet of 
ordinary paper; 
in contrast to a rubber sheet, 
expanding or contracting a sheet of paper 
is hard, although bending is easy. 

\begin{figure}[t]
\begin{center}
\includegraphics[width=8cm,clip]{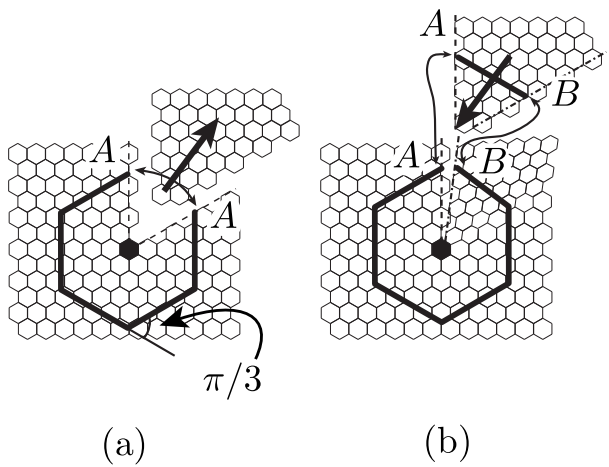}
\end{center}
    \caption{
	(a) Formation of a $\pi/3$ wedge disclination and (b)
	a $-\pi/3$ wedge disclination. 
	(a) Geodesic polygon around a $\pi/3$ 
	disclination, and (b) that around a $-\pi/3$ 
	disclination are also shown with bold lines.  
	The points marked with $A$'s and $B$'s should be merged 
	in each figure, 
	so that the bold lines are closed. 
    }
    \label{disc1}
\end{figure}

The Gaussian curvature is generated 
in the graphitic network only by introducing disclinations. 
Disclination is a topological defect 
created in the following way: 
At the center of a hexagon of a graphite 
sheet, there is a six-fold symmetry. 
Now we cut off one of the six equivalent 
fan-shaped regions as shown in 
Fig. \ref{disc1}(a) and join the two 
edges created by cutting. 
This operation leaves a pentagon 
at the center. 
The structure obtained in this way 
is called $\pi/3$ wedge disclination. 
If we cut off two of the fan-shaped 
regions, we obtain $2 \pi/3$ wedge disclination. 
In this case, there is a square at the center. 
Usually a $2 \pi/3$ wedge disclination 
has much higher energy than a 
$\pi/3$ wedge disclination and 
is almost negligible. 
(Instead of a $2 \pi/3$ wedge disclination, 
we may introduce two $\pi/3$ ones 
at some distance from each other.)
On the other hand, if we add one 
extra fan-shaped region as shown in 
Fig. \ref{disc1}(b), the obtained 
structure is $-\pi/3$ wedge disclination, 
which has a heptagon at the center. 
In this case also, octagons are less 
favored than heptagons. 
From now on we use the word 
\lq\lq disclination\rq\rq instead of 
\lq\lq wedge disclination\rq\rq for short. 

Next we calculate the Gaussian 
curvature generated by a disclination. 
Here we consider only one disclination 
in a graphite sheet. 
The Gaussian curvature of the 
disclination is calculated from 
the Gauss-Bonnet theorem, 
\begin{align}
\iint_{D} K {\rm d}A =2 \pi 
- \sum_{i=1}^{N} \theta_i, 
\label{GB_1}
\end{align}
where $D$ is the region surrounded 
by a geodesic polygon 
on the surface (graphite sheet), 
which is shown in Fig. \ref{disc1}
by bold lines, 
and the l.h.s. 
is the surface 
integral of the Gaussian curvature 
$K$ over the region $D$. 
In the present case, $K$ is non-zero 
only at the center where the disclination 
is located. 
$\theta_i$ is the angle between the 
neighboring sides of the polygon 
(exterior angles) at the $i$-th vertex 
(See Appendix \ref{app_GB_1} for details.)
We can easily see from  Fig. \ref{disc1} (a) 
that the r.h.s. of Eq. (\ref{GB_1}) reads 
$2 \pi - 5 \pi/3 = \pi/3$. 
Therefore the Gaussian curvature 
generated by a $\pi/3$ disclination is 
$\pi/3$. 
In the same way, the Gaussian curvature 
generated by a $-\pi/3$ disclination is 
$-\pi/3$, by considering the geodesic 
polygon as shown in Fig. \ref{disc1} (b). 

\section{Gauss-Bonnet theorem and its 
application to various 
graphitic carbon materials}
\label{G-B_application}

\subsection{Fullerenes}
\label{fullerene}

\begin{figure}[t]
\begin{center}
\includegraphics[width=6cm,clip]{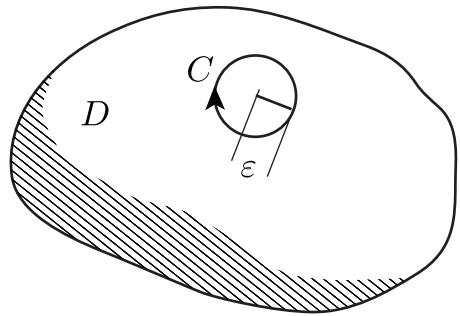}
\end{center}
    \caption{
	A simply connected closed surface. 
	$C$ is a small circle on the 
	surface and $D$ is the region 
	outside of the circle $C$. 
    }
    \label{sphere}
\end{figure}

Fullerenes are simply connected closed 
surface formed by a graphite sheet. 
Here we calculate the total curvature of 
the closed surface using the Gauss-Bonnet theorem. 
We consider the surface as depicted 
in Fig. \ref{sphere} and 
draw a small circle $C$ with radius 
$\varepsilon$ on it. 
The circle is so small that the surface in the circle 
can be regarded as a plane. 
The region $D$ in this case is the outside 
of the small circle. 

We apply the Gauss-Bonnet theorem, 
\begin{align}
\int_{C} \kappa_g {\rm d}s +  
\iint_{D} K {\rm d}A =2 \pi,
\label{GB_sp}
\end{align}
where $\kappa_g$ is the geodesic curvature of $C$ 
(see Eq. (\ref{GBeq_1}) of Appendix \ref{app_GB}).
The geodesic curvature of the circle coincides with the 
curvature of the circle, as one can see 
from the definition of the geodesic curvature 
given in Appendix \ref{app_GB_1}, 
and we obtain $\kappa_g=-1/\varepsilon$ 
(Note that the minus sign comes from the 
fact that the inside of the circle 
is the outside of the region $D$). 
Then the first term of Eq. (\ref{GB_sp}) 
equals $2 \pi \varepsilon \times (-1/\varepsilon) = - 2 \pi$ 
and 
we obtain 
\begin{align}
\iint_{D} K {\rm d}A = 4 \pi. 
\end{align}
This result does not depend on the 
details of the surface if the surface is a simply connected 
closed one. 
In the limit of $\varepsilon \rightarrow 0$, $D$ 
covers the whole region of the surface. 

Since the Gaussian 
curvature per a $\pi/3$ disclination is 
$\pi/3$, twelve $\pi/3$ disclinations are required to 
make a graphite sheet closed. 
This means that there are always twelve 
pentagons on a simply connected 
closed graphitic carbon material, 
which is a well-known result for Fullerenes.

\subsection{Cones}
\label{cone}

\begin{figure}[t]
\begin{center}
\includegraphics[width=6cm,clip]{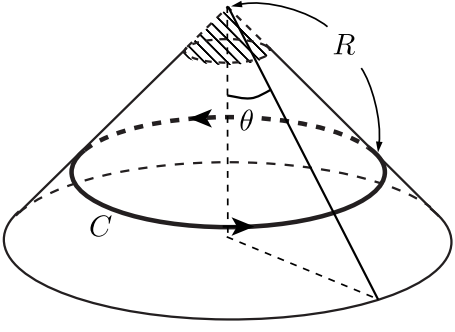}
\end{center}
    \caption{
A circular cone. Disclinations are 
located in the shaded region. 
$C$ is a circle on the cone whose 
distance from the vertex is $R$. 
$\theta$ is the angle between the 
generating line and the center line of the 
cone. 
    }
    \label{cone_fig}
\end{figure}

Next we consider the circular 
cone-shaped graphitic carbon material 
as depicted in Fig. \ref{cone_fig}. 
Here we consider a circle $C$ on 
the cone as shown in Fig. \ref{cone_fig}, 
where the distance from the vertex of 
the cone to the circle is $R$. 
The geodesic curvature of $C$ is 
estimated by developing the side 
of the cone on a plane. 
The curve $C$, when developed on a plane, 
becomes a circle of radius $R$. 
Therefore its curvature, $1/R$, gives the 
geodesic curvature of $C$ on the cone. 
(Note that the geodesic curvature of a curve 
on a developable surface is obtained by 
developing the surface on a plane and 
measuring the curvature of the curve.) 
Integrating the geodesic curvature along 
the circle, we obtain 
$2 \pi R \sin \theta \times (1/R)$, 
where $\theta$ is the angle between 
the generating line and the center line of the cone. 
From Eq. (\ref{GB_sp}) we obtain the 
Gaussian curvature inside of the curve $C$ 
as 
\begin{align}
\iint_{D} K {\rm d}A = 2 \pi (1 - \sin\theta). 
\end{align}
This curvature is generated by the 
disclinations existing in the shaded 
region of the cone 
in Fig. \ref{cone_fig}. 
Therefore it should be quantized in 
the unit of $\pi/3$. 
If we assume that the total number 
of disclinations 
(the number of $\pi/3$ disclinations 
minus that of $-\pi/3$ disclinations) 
is $n$, we obtain $\theta = \theta_n$, 
where 
\begin{align}
\theta_n=\arcsin \left(1 - \frac{n}{6}\right). 
\end{align}
Here we assume that $n$ is positive 
(see Discussion for the case of negative $n$). 
In order that $0 < \theta_n < \pi/2$, 
$n$ is limited as $1 \leq n \leq 5$ and 
we obtain $\theta_1=56.4^\circ$, 
$\theta_2=41.8^\circ$, 
$\theta_3=30^\circ$, $\theta_4=19.5^\circ$ 
and $\theta_5=9.6^\circ$
(in degree). 
In case of $n=6$, we have a tube 
in the limit of an infinitely sharp cone. 

The results obtained until now are 
rather trivial ones, 
which are also obtained in terms of 
more elementary arguments. 
In the next section we see a little 
more non-trivial cases. 

\subsection{Kinked nanotube}
\label{kinked_tube}

\begin{figure}[t]
\begin{center}
\includegraphics[width=8cm,clip]{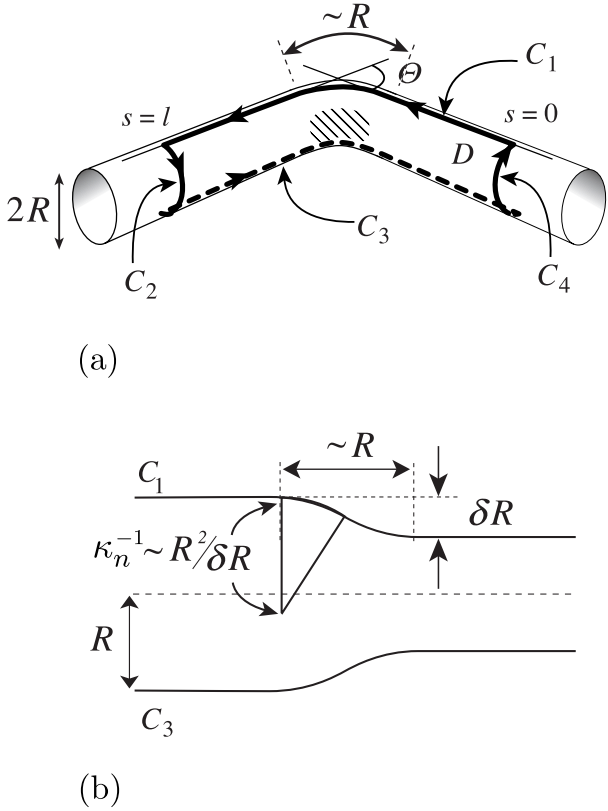}
\end{center}
    \caption{
(a) A kinked tube. Some negative 
disclinations are expected in the shaded region and 
the same number of positive ones on the other side. 
Except for the vicinity of the disclinations, 
the surface is assumed to be free from 
Gaussian curvature, {\it i.e.}, $K=0$. 
$D$ is the region on the surface surrounded 
by $C=C_1+C_2+C_3+C_4$. 
$C_2$ and $C_4$ are perpendicular 
to the axis of the tube. 
$C_1$ ($C_3$) is the line on the upper 
(lower) part of the tube. 
In this figure the tube does not change 
its radius at the kink position. 
(b) Cross section of a tube 
in case that the tube changes 
its radius near the kink position. 
The cross section is along the curves $C_1$ and $C_3$. 
The radius of the curvature 
near the kink is estimated to be 
$\sim R^2/\delta R$ from this figure, 
which corresponds approximately to the normal 
curvature $\kappa_n$ of the curve $C_1$ and 
$C_3$. 
}
    \label{kink}
\end{figure}

Here we consider a kinked nanotube. 
If there are several $\pi/3$ 
disclinations on one side of a tube 
and the same number of $-\pi/3$ disclinations on the 
other side, 
the tube has a kink structure, 
as depicted in Fig. \ref{kink}(a). 
We denote the corner angle of the kink by $\Theta$. 
We assume that the negative (positive) 
disclinations are located in the shaded region 
(the region behind the tube not depicted here) and 
except for these regions the surface 
is free from the Gaussian curvature ($K=0$) 
\cite{kink}. 
Now we consider a curve $C=C_1+C_2+C_3+C_4$ 
in Fig. \ref{kink}(a). 
The region surrounded by $C$ is denoted by $D$. 
$C_2$ and $C_4$ are perpendicular to the 
axis of the tube and comprise
geodesic curves. 

Next we look at the curves $C_1$ and $C_3$. 
Here we assume a very simple situation: 
$C_1$ and $C_3$ are the top line and the 
bottom line of the 
tube, respectively and are plane curves, 
namely each curve is 
included in a single plane. 
The planes are the tangent planes of the 
tube. 
This can be applied to the case that the 
tube does not change 
its radius at the kink. 
The deviation from this situation is also 
discussed later in this section. 

Under the above assumption, we now apply 
the Gauss-Bonnet theorem to 
the kinked tube. 
Since $C_1$ and $C_3$ lie in a tangent 
plane of the tube, 
at any points on these curves the normal 
curvature $\kappa_n$ vanishes. 
Therefore the curvature of the curve 
coincides with the 
geodesic curvature. 
(See Appendix \ref{app_GB_1}.) 

Then we apply the Gauss-Bonnet theorem to 
the curve $C$ and the region $D$. 
Here we use the following form
\begin{align}
\int_{C'} \kappa_g {\rm d}s +  \iint_{D} K {\rm d}A =2 \pi 
- \sum_{i=1}^{N} \theta_i, 
\label{GB_kink}
\end{align}
(see Appendix \ref{app_GB_2}). 
$C'$ means that the integral of 
$\kappa_g$ is preformed except 
for the vertex points. 
Here $\theta_i$ is the exterior angle at 
the vertex $i$ of the line $C$. 
On $C_2$ and $C_4$ the first term vanishes 
since they are geodesic curves. 
On $C_1$ and $C_3$ the geodesic 
curvature coincides with the curvature 
of the curves. 
Since $C_1$ and $C_3$ are plane curves, 
their curvature $\kappa$ is given by 
$\kappa = {\rm d} \theta(s) /{\rm d} s$, 
where $\theta (s) $ is the angle between 
the tangent line of the curve and a certain 
standard line on the plane
with $s$ being the length measured along the line. 
We set the sign of $\theta(s)$ so that 
it increases when the curve is convex 
as viewed from the inside of the region $D$. 

Let us calculate the geodesic curvature of $C_1$. 
Using the parameter $s$ along $C_1$, 
the curvature of $C_1$ is 
given by ${\rm d} \theta /{\rm d} s$ and 
it gives the geodesic curvature of $C_1$. 
Therefore we can evaluate the first term 
of Eq. (\ref{GB_kink}) as 
\begin{align}
\int_{C_1} \kappa_g {\rm d}s = \int_0^l 
\frac{{\rm d} \theta}{{\rm d} s} {\rm d}s
=\theta(l) - \theta(0) = \Theta, 
\label{int_kappa_g}
\end{align}
where $l$ is the length of $C_1$ and 
$s=0$ corresponds to the starting point of $C_1$. 
$C_3$ also gives the same contribution. 

Noting that the exterior angles between 
$C_1$ and $C_2$, $C_2$ and $C_3$, 
$C_3$ and $C_4$, and $C_4$ and $C_1$ are 
all $\pi/2$, we can calculate the 
Gaussian curvature as 
\begin{align}
 \iint_{D} K {\rm d}A =  - 2 \Theta. 
\end{align}

If we set $D$ so that it covers the outer 
half of the tube (opposite side with the above $D$), 
we have the result 
\begin{align}
 \iint_{D} K {\rm d}A =  2 \Theta. 
\end{align}

Since the Gaussian curvature is quantized 
in the unit of $\pi/3$ as is mentioned before, 
$\Theta$ can take only the discrete values, 
multiples of $\pi/6$. 
This is the most important result 
of this paper. 

Now we consider what happens if the 
curves $C_1$ and $C_3$ are 
not plane curves. 
For example, when the tube changes 
its radius at the kink, 
this situation appears. 
In this case, however, there is no 
simple way to calculate the first 
term of Eq. (\ref{GB_kink}). 
Therefore we give a rough estimation 
of the deviation 
from the result obtained in Eq. (\ref{int_kappa_g}). 

We consider the case that the tube 
changes its radius from $R$ to $R + \delta R$ 
near the kink position. 
It is expected that the length scale 
needed for this change is probably 
of the order of $R$, which is 
also the size of the kink region 
(the transition region from one 
straight tube to the other). 
Therefore in this case we expect that 
$C_1$ changes as shown in Fig. \ref{kink}(b), 
in which the cross section of the tube 
along the curves, $C_1$ and $C_3$, is shown. 
Assuming that $C_1$ and $C_3$ interpolate 
smoothly between two 
parallel lines outside of the kink, 
as shown in Fig. \ref{kink}(b), 
let us estimate the change of 
the first term of Eq. (\ref{GB_kink}). 
Here we limit ourselves to the case of $C_1$. 
The case of $C_3$ is similar. 
First we rewrite $\kappa_g$ 
in terms of the normal curvature $\kappa_n$ and 
the curvature of the curve $\kappa$ (see Appendix 
\ref{app_GB_1} 
for the definitions of $\kappa_n$), 
\begin{align}
\kappa_g &=\sqrt{\kappa^2 - \kappa_n^2}. 
\nonumber
\end{align}
(We assumed that $\kappa_g$ has a definite sign 
in the kink region.)
Note that $\kappa_g$ and $\kappa$ are 
the same order $\sim \Theta/R$, 
since the length of the kink region is 
roughly $\sim R$ as one can see from 
Fig. \ref{kink}(a), and the line integral of $\kappa$ 
along the kink region is of the order of $\Theta$. 
On the other hand, $\kappa_n$ is much smaller. 
We estimate $\kappa_n$ 
generated by the radius change to be 
of the order of $\sim \delta R/R^2$ 
as seen in Fig. \ref{kink}(b). 
We now obtain 
\begin{align}
\int_{C_1} \kappa_g ds &=
\int_{C_1} \sqrt{\kappa^2 - \kappa_n^2}ds 
\nonumber\\
&\simeq \Theta
+  {\cal O}\left(
\frac{\delta R}{R}\right)^2.
\end{align}
We see that, even if the change of 
the radius at the kink is about 10\%, 
the correction to the present 
evaluation is a few percent, 
which is not too large to spoil the 
quantization of $\Theta$ \cite{kappa}. 

\subsection{Y-junction and other forms with branches}

\begin{figure}[t]
\begin{center}
\includegraphics[width=8cm,clip]{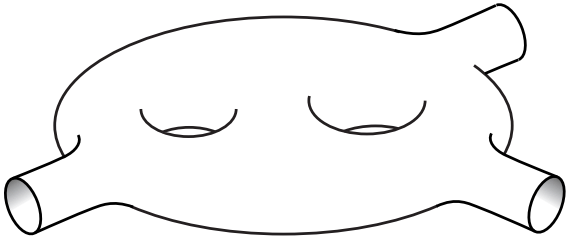}
\end{center}
    \caption{
	Graphitic carbon material with several 
	branches of nanotubes. 
	The important feature is the 
	number of the branches (3 in this case)
	and the genus of the 
	\lq\lq body\rq\rq (2 in this case). 
}
    \label{branch}
\end{figure}

Recently more intriguing structures of 
graphitic carbon materials are created, including 
Y-junctions\cite{Andriotis,Martel,Nardelli,
Shibata,Terrones,Peng,Zhao}. 
Here we describe a general properties 
derived from the Gauss-Bonnet theorem. 
Let us consider a graphitic carbon material  
depicted in Fig. \ref{branch}. 
The details of the structure 
is not important. 
We denote the number of the 
branches (tubes) going out of the structure by $N$. 

We note that we can make this structure 
a closed one by putting 
half of spherical Fullerenes 
(caps) to all the ends of the branches. 
In that case, the number of the 
disclinations is increased by $6 N$, 
since there are six in each cap. 

Here we employ a more general 
form of the Gauss-Bonnet theorem 
for closed surfaces 
(see Appendix \ref{app_GB_2} for more details), 
\begin{align}
\iint_S K {\rm d}A = 4 \pi (1 - g), 
\label{GB_multi}
\end{align}
where $g$ is the genus of the surface, 
which is zero for a simply connected 
closed surface like a sphere, 
and one for a torus. 
In case of Fig. \ref{branch}, 
$g$ equals two. 

Now we can estimate the number of 
disclinations in the original structure without caps. 
Since the integrated gaussian 
curvature of the capped surface is $4 \pi (1 - g)$, there 
should be $12 (1 - g)$ disclinations in total. 
Among them, $6 N$ disclinations 
come from the caps. 
Therefore the number of disclinations 
in the original structure is 
$12 (1 - g) - 6 N$. 

In case of a Y-junction, $g=0$ and $N=3$, 
and we see that the total number of disclinations 
is $-6$. This means that there are 
six $-\pi/3$ disclinations, 
which is consistent with the 
structure of graphitic carbon material given 
in Ref. \citen{Andriotis}. 

\section{Discussion and Further Problems}
\label{Discussion}

The most non-trivial result in this paper 
is that the corner angle of 
the kinked nanotube is quantized by $\pi/6$. 
Although there can be a small deviation in real systems, 
it is estimated to be small. 
Now we discuss how this nature 
of the nanotube is observed in 
actual experiments. 
The forms of nanotubes are 
observed by using the transmission electron microscope. 
However, in actually synthesized 
materials, nanotubes are quite 
randomly piled as seen in 
Ref. \citen{Iijima_nanotube}. 
In such a situation, each 
tube undergoes stress from other 
tubes and is elastically deformed. 
Since such deformation is not taken into 
account in our study, the prediction of 
this paper does not 
seem to be easily 
realized in realistic materials. 
In order to overcome this 
difficulty one may 
try the following two methods. 
Firstly, if a material can be well separated 
from other materials and examined 
individually, the prediction 
in this paper is easily ascertained. 
This kind of methods may be done by using the 
developed nanotechnology in the future. 
Secondly, one may adopt statistical way to 
examine the forms of carbon materials. 
Since the most stable forms of graphitic carbon materials 
are those predicted in this paper, 
these forms are more easily realized than the 
elastically deformed ones. 
Therefore, in case of the kinked tubes, for example, 
if one gather data of the kink angles 
for many kinked tubes and plot the number of 
kinks as a function of the kink angle, 
the distribution shows peaks at the 
angles predicted in this paper, namely 
$\pi/6,\pi/3,\pi/2 \cdots$. 
In any way, to measure the forms of the 
carbon materials precisely, further developments 
may be needed in both hardware and software 
of the experimental technique. 
For example, to measure the kink angle of a 
kinked tube, one has to know the three dimensional 
configuration of the tube. 
It does not seem to be easily 
carried out by the present 
electron microscope technology, 
however it is not impossible in the future. 

Next we discuss more theoretical aspect of the argument of 
this paper. 
It should be noted that we made some implicit assumption 
on the energetics of the graphitic carbon materials. 
For example, in Sec. \ref{cone} we have assumed that 
the cones are not elliptic cones but 
circular cones. 
This is because the latter seems to be more 
stable than the former from a natural intuition. 
We have also made similar assumptions in 
other arguments of this paper. 
If we argue these points more precisely, 
we have to calculate the elastic energy of the 
carbon materials. 
To do this we need to introduce a quantity 
which is not treated in this paper, 
that is the so-called {\it mean curvature} $H$ 
(see Appendix \ref{Gaussian_curvature}). 
As seen from the 
famous Gauss's \lq\lq {\it Theorem 
egregium} ({\it a most excellent theorem})\rq\rq, 
the Gaussian curvature is 
related only to the intrinsic 
properties of a surface. 
The word \lq\lq intrinsic properties\rq\rq means 
the properties that can be 
reached by measuring the 
lengths between arbitrary two points on 
the surface, namely, without looking 
from the outside of the surface. 
For a simple example, human beings 
before Gagarin  
could confirm the fact 
that the earth is round by just 
navigating all over the world. 
However, if the earth is cone-shaped, 
it is absolutely impossible to say whether the 
cone is a circular one or an elliptic one 
by only navigating on it, 
since they have the same 
intrinsic properties. 
This information can be 
obtained only by looking the surface 
from the outside. 
Looking from the outside is, actually, 
equivalent with measuring the mean curvature. 
Therefore the elastic energy of 
a carbon material should be 
related to the mean curvature. 
Some more efforts are needed to construct the 
theory which treats this problem. 
However we leave it for future studies. 
Here is one unsolved problem: 
\begin{enumerate}
\item[]
If there is a disclination with positive curvature 
in a graphite sheet, 
the surface forms a circular cone. 
Then what if the curvature is negative? 
\end{enumerate}
The answer is not trivial but the situation 
like this may be possible in 
actual graphitic carbon materials. 
I hope someday such intriguing graphitic 
carbon materials are actually observed. 

\section{Summary}
\label{Summary}

In this paper, we have applied 
the knowledge of differential geometry 
to the morphology of graphitic 
carbon materials. 
We have clarified 
how the quantization of the Gaussian curvature 
affects the forms of the graphitic carbon materials, 
such as spheres, cones, kinked tubes and structures 
with branches including Y-junctions. 
Especially it is predicted that the corner 
angle of a kinked nanotube is quantized by $\pi/6$. 
Experimental methods to observe these properties 
are also discussed. 

\appendix
\section{Curvature}
\label{curvature}

\subsection{Curvature of a curve}

\begin{figure}[t]
\begin{center}
\includegraphics[width=6cm,clip]{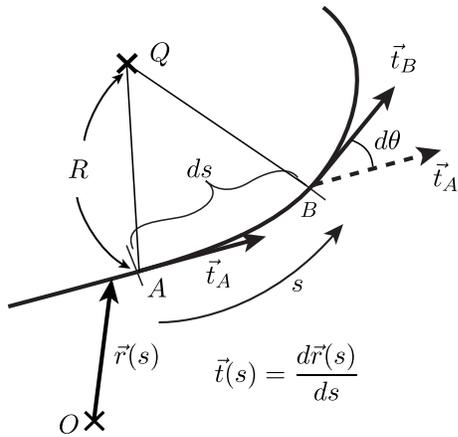}
\end{center}    
\caption{
Curve $C$, parametrized by the length $s$, 
and its tangent vectors 
$\vec{t}_A$ and $\vec{t}_B$ 
at points $A$ and $B$, respectively. 
The distance between $A$ and $B$ 
is infinitesimal ${\rm d} s$. 
The angle between $\vec{t}_A$ 
and $\vec{t}_B$ 
defines ${\rm d}\theta$ and 
$|{\rm d}\theta/{\rm d}s|$ 
gives the curvature $\kappa$ of 
$C$ at $A$ in the limit of 
${\rm d} s\rightarrow 0$. 
The center of the curvature 
is denoted by $Q$ and 
$R$ is the radius of the curvature. 
    }
    \label{curve}
\end{figure}

First we begin with the 
definition of the curvature of a curve. 
As shown in Fig. \ref{curve} we parametrize curve 
$C$ by a parameter $s$ which is 
the length along $C$. 
Points on $C$ is then written as $\vec{r}(s)$. 
Now consider two points on the curve indicated 
by $A$ and $B$, the distance between 
which is ${\rm d} s$ (an infinitesimal number).  
The tangent vectors, 
given by ${\rm d} \vec{r}/{\rm d} s$, 
at $A$ and $B$ are
denoted by $\vec{t}_A$ and 
$\vec{t}_B$, respectively. 
We define ${\rm d}\theta$ as the angle between 
$\vec{t}_A$ and $\vec{t}_B$. 
Now the curvature $\kappa$ of $C$ is given by 
\begin{align}
\kappa = \left|\frac{{\rm d} \theta}{{\rm d} s}\right| = 
\left|\frac{{\rm d}^2 
\vec{r}}{{\rm d} s^2}\right|. 
\label{curvature}
\end{align}
The center of curvature, 
denoted by $Q$ in Fig. \ref{curve}, 
is obtained by considering 
the curve $A$$B$ as a 
part of a circle and finding 
the center of it. 
The radius of the circle $R$ is called the 
radius of curvature and 
is related to the curvature $\kappa$ 
by $\kappa=1/R$.  

In case of a plane curve, 
which is a curve included in a single plane, 
we can define $\kappa$ by 
${\rm d} \theta/{\rm d} s$ 
with a sign by appropriately 
choosing the direction $\theta=0$. 
In this case, the integral of 
$\kappa$ from a point $A$ ($s=s_A$)
to $B$ ($s=s_B$) gives the 
angle made by the tangent lines of $C$ 
at $A$ and $B$, 
\begin{align}
\int_{s_A}^{s_B} {\rm d}s\,\,\kappa=
\int_{s_A}^{s_B} {\rm d}s\,\,
\frac{{\rm d} \theta}{{\rm d}s}
=\theta(s_B) - \theta(s_A). 
\label{plane_curvature}
\end{align}

\subsection{Gaussian curvature of a surface}
\label{Gaussian_curvature}

\begin{figure}[t]
\begin{center}
\includegraphics[width=8cm,clip]{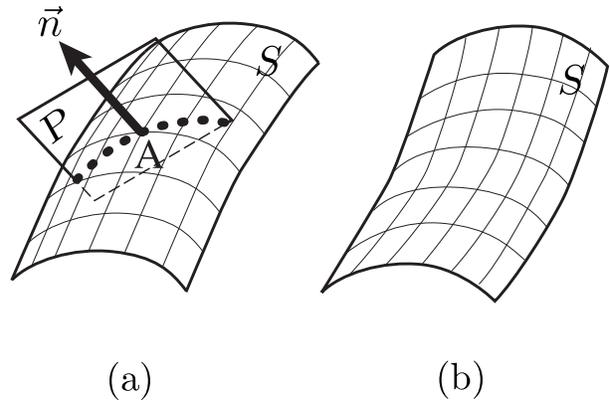}
\end{center}
    \caption{
	(a) Surface $S$, its normal vector $\vec{n}$ 
	at point $A$ and a plane $P$ including 
	$\vec{n}$. Cross section between $A$ and $P$ 
	is shown by a dotted line. 
	(b) A surface with negative Gaussian curvature. 
    }
    \label{surface}
\end{figure}

Next we describe the Gaussian 
curvature of a curved surface. 
As shown in Fig. \ref{surface}(a), 
we consider the normal vector $\vec{n}$ of the surface $S$ at 
point $A$. 
Then consider a plane $P$ which 
includes $\vec{n}$. 
The cross section of $S$ and $P$ 
defines a plane curve (dotted line). 
The curvature of this curve depends 
on the direction of $P$. 
If we rotate $P$ around $\vec{n}$, 
the curvature changes and 
takes a maximum and a minimum before 
we rotate $P$ by $\pi$. 
(Note that the curvature can be either 
positive or negative.)
Let $\kappa_1$ and $\kappa_2$ 
be the maximum and the minimum of the 
curvature, respectively. 
Then the Gaussian curvature $K$ at $A$ is 
given by the product $\kappa_1 \kappa_2$. 
Another curvature frequently 
used in the differential 
geometry is 
the mean curvature $H$ defined 
by $(\kappa_1 + \kappa_2)/2$. 
At so-called {\it navel points} 
we find $\kappa_1=\kappa_2$. 

The Gaussian curvature can be either positive 
or negative. 
Surfaces with positive and negative Gaussian 
curvature are depicted in 
Fig. \ref{surface}(a) and (b), respectively. 

A surface with vanishing Gaussian curvature is 
called a {\it developable surface}. 
If we cut off a piece of a developable surface, 
we can develop the piece on a plane without 
expanding or contracting any part of 
the piece. 
For example, sides of cylinders and cones are 
developable surfaces, 
but spheres are not. 

\appendix
\section{Gauss-Bonnet theorem}
\label{app_GB}

\subsection{Geodesic curvature}
\label{app_GB_1}

\begin{figure}[t]
\begin{center}
\includegraphics[width=8cm,clip]{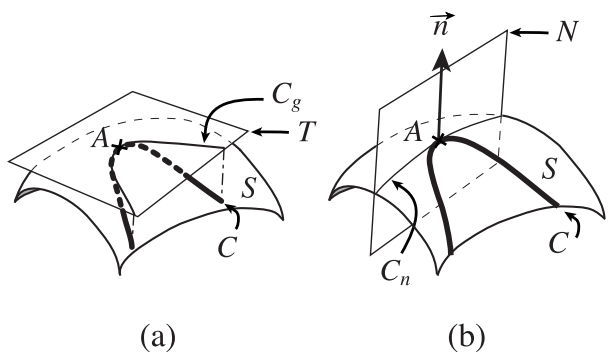}
\end{center}
    \caption{
	$C$ is a curve on a surface $S$. 
	(a) The plane denoted by $T$ is 
	the tangent plane of $S$ at a point $A$, where 
	$A$ is on $C$. 
	The orthogonal projection of $C$ onto 
	$T$ is $C_g$. 
	The curvature of $C_g$ at $A$ gives 
	geodesic curvature $\kappa_g$ 
	of $C$ at $A$. 
	(b) $N$ is the normal plane of $S$ at $A$, 
	which is parallel to the normal 
	vector $\vec{n}$ of $S$ at $A$. 
	The cross section of $S$ and 
	$N$ is a curve denoted by $C_n$. 
	The curvature of $C_n$ at $A$ gives the normal 
	curvature $\kappa_n$ 
	of $C$ at $A$. 
	Note that the direction of 
	$N$ is determined so that $C_n$ 
	and $C$ have a common tangent line at $A$. 
    }
    \label{geodesic}
\end{figure}

Before introducing the Gauss-Bonnet theorem, 
we describe the definition of the 
geodesic curvature and geodesic curves. 
Geodesic curvature is defined for a curve 
on a surface. 
In Fig. \ref{geodesic}, we have shown a curve 
$C$ on a surface $S$. 
Now consider a tangent plane of $S$ at point 
$A$, which we denote by $T$ 
in Fig. \ref{geodesic} (a). 
The point $A$ is also on the curve $C$. 
Then consider the orthogonal projection of $C$ onto $T$, 
which we denote by $C_g$. 
Geodesic curvature of the curve $C$ at 
$A$ is given by the 
curvature of the curve $C_g$ at $A$. 
Let us denote it by $\kappa_g$. 
A curve on a surface whose geodesic curvature 
vanishes is called a geodesic curve. 
The sign of $\kappa_g$ is determined by 
introducing the orientation of $S$. 
This point will be discussed later in relation to the 
Gauss-Bonnet theorem. 

Next we consider a normal vector $\vec{n}$ of 
the surface $S$ at $A$ and, 
then, consider a plane $N$ which includes 
$\vec{n}$. 
The cross section between $S$ and $N$ 
defines a curve, which we denote by $C_n$. 
Here we set the direction of $N$
so that $C_n$ and $C$ have 
a common tangent line at $A$.  
The curvature of $C_n$ at $A$ gives the 
normal curvature of $C$ at $A$, 
which we denote by $\kappa_n$. 
Let the curvature of $C$ at $A$ 
be denoted by $\kappa$, 
then the following relation holds, 
\begin{align}
\kappa^2 = \kappa_n^2 + \kappa_g^2. 
\label{curvature_sum}
\end{align}

\subsection{Gauss-Bonnet theorem}
\label{app_GB_2}

The Gauss-Bonnet theorem is expressed in several forms. 

\begin{figure}[t]
\begin{center}
\includegraphics[width=5cm,clip]{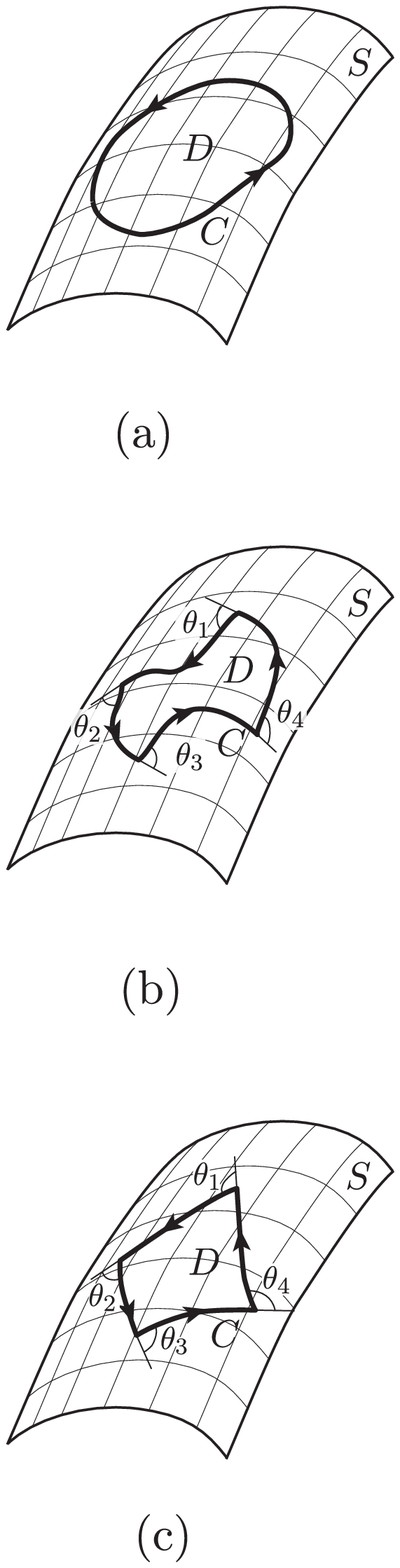}
\end{center}
    \caption{
(a) A smooth curve $C$, 
(b) a sectionally smooth curve 
$C$ with several vertices, and 
(c) a polygon $C$ whose sides are geodesic 
curves on the surface $S$. 
$D$ is the region surrounded by $C$. }
    \label{GB}
\end{figure}

First we consider simply 
connected region $D$  on a surface $S$ 
surrounded by a smooth curve, 
denoted by $C$, 
as shown in Fig. \ref{GB} (a). 
The Gauss-Bonnet theorem is expressed as 
\begin{align}
\int_{C} \kappa_g {\rm d}s +  
\iint_{D} K {\rm d}A =2 \pi, 
\label{GBeq_1}
\end{align}
where $\kappa_g$ is the geodesic 
curvature of $C$, 
and the second term is the 
integration of the Gaussian 
curvature $K$ over $D$. 
Here the geodesic curvature 
is defined with a sign: 
we set $\kappa_g > 0$ ($\kappa_g < 0$)
if the curve is 
convex (concave) 
viewing from the inside of $D$. 
This equation relates the total 
Gaussian curvature in the region $D$ and 
the geodesic curvature of the 
boundary $C$. 

If we consider a sectionally smooth 
curve with $N$ 
vertices as shown in Fig. \ref{GB}(b), 
the Gauss-Bonnet theorem reads, 
\begin{align}
\int_{C'} \kappa_g {\rm d}s +  
\iint_{D} K {\rm d}A =2 \pi 
- \sum_{i=1}^{N} \theta_i, 
\label{GBeq_2}
\end{align}
where $C'$ means that the integral is 
evaluated except for the vertex points. 
$\theta_i$ is the exterior 
angles of the vertices, 
as shown in Fig. \ref{GB}(b). 

If we consider a geodesic 
polygon, which is a 
polygon with all the sides begin 
geodesic curves, the 
Eq. (\ref{GBeq_2}) is 
further reduced to 
\begin{align}
\iint_{D} K {\rm d}A =2 \pi 
- \sum_{i=1}^{N} \theta_i, 
\label{GBeq_3}
\end{align}
since the geodesic curvature 
of the boundary vanishes.

If $S$ is not a simply connected surface, 
the Gauss-Bonnet theorem has a more 
profound form. 
Here we limit ourselves 
to the case of the closed surfaces. 
Now, $D$ covers the 
whole region of $S$. 
We obtain the following theorem, 
\begin{align}
\iint_S K {\rm d}A = 2 \pi \chi(S) = 4 \pi (1 - g), 
\label{GB_multi}
\end{align}
where $\chi(S)$ is the Euler characteristic 
of the surface, 
given by $\chi(S) = F - E + V$ where $F$, 
$E$ and $V$ are the number of 
faces, edges and vertices that occur in the 
triangulation of the surface. 
Here $g$ is the genus of the surface, {\it e.g.}, 
zero for a sphere, one for a torus. 
See, for example, Ref. \citen{Struik} 
for more details.


\begin{thebibliography}{99}
\bibitem{Kroto} 
H. W. Kroto, J. R. Heath, S. C. O'Brien, 
R. F. Curl and R. E. Smalley, 
Nature {\bf 318} (1985) 162. 
\bibitem{Osawa}
E. Osawa, Kagaku {\bf 25} (1970) 854 (in Japanese). 
\bibitem{Iijima_C60}
S. Iijima, J. Phys. Chem. {\bf 91} (1987) 3466. 
\bibitem{Curl-Smally}
R. F. Curl and R. E. Smally, Science {\bf 242} (1988) 1017. 
\bibitem{Kroto_Review}
H. W. Kroto, Science {\bf 242} (1988) 1139.
\bibitem{Iijima_nanotube}
S. Iijima, Nature {\bf 354} (1991) 56. 
\bibitem{Kroto_Cn}
H. W. Kroto, Nature {\bf 329} (1987) 529. 
\bibitem{Kratschmer}
W. Kr\"atschmer, L. D. Lamb, 
K. Fostiropoulos and D. R. Huffman, 
Nature {\bf 347} (1990) 354. 
\bibitem{Iijima_curvature}
S. Iijima, T. Ichihashi and Y. Ando, Nature {\bf 356} (1992) 776. 
\bibitem{Iijima_single}
S. Iijima and T. Ichihashi, Nature {\bf 363} (1993) 603. 
\bibitem{Yao}
Z. Yao, H. W. Ch. Postma, L. Balents and C. Dekker, 
Nature {\bf 402} (1999) 273. 
\bibitem{Satishkumar}
B. C. Satishkumar, P. J. Thomas, 
A. Govindaraj and C. N. R. Rao, 
Appl. Phys. Lett. {\bf 77} (2000) 2530. 
\bibitem{Kroto-McKay}
H. W. Kroto and K. McKay, Nature {\bf 331} (1988) 328. 
\bibitem{McKay-Terrones}
A. L. McKay and H. Terrones, Nature {\bf 352} (1991) 762. 
\bibitem{Lenosky}
T. Lenosky, X. Gonze, M. Teter 
and V. Elser, Nature {\bf 355} (1992) 333.
\bibitem{MoS2}
G. Seifert, H. Terrones, M. Terrones, 
G. Jungnickel and T. Frauenheim, 
Phys. Rev. Lett. {\bf 85} (2000) 146. 
\bibitem{Tanda_private}
S. Tanda, private communication. 
\bibitem{Tanda_NbSe3}
S. Tanda, T. Tsuneta, Y. Okajima, K. Inagaki, 
K. Yamaya and N. Hatakenaka, Nature {\bf 417} (2002) 397. 
\bibitem{kink}In case of kinked tubes, it is not clear whether 
the region outside of the disclinations is really free 
from Gaussian curvature. However, for sufficiently large tubes 
as compared to the hexagons of the graphite sheet, 
it can be assumed that the Gaussian curvature is 
vanishingly small apart from the disclinations. 
\bibitem{kappa}
There may be some correction arising from the change of $\kappa$. 
We found that this is the same order as that from $\kappa_n$. 
This point will be discussed elsewhere. 
\bibitem{Andriotis} A. N. Andriotis, 
M. Menon, D. Srivastava and L. Chernozatonskii, 
Phys. Rev. Lett. {\bf 87} (2001) 066802-1. 
\bibitem{Martel} R. Martel, H. R. Shea and P. Avouris, 
Phys. Chem. {\bf 36} (1999) 7551. 
\bibitem{Nardelli} M. B. Nardelli and J. Bernholc, 
Phys. Rev. B {\bf 60} (1999) R16 338. 
\bibitem{Shibata}
Y. Shibata and S. Maruyama, Physica B{\bf 323} (2002) 
187. 
\bibitem{Terrones} 
M. Terrones, F. Banhart, N. Grobert, J.-C. Charlier, H. Terrones 
and P. M. Ajayan, Phys. Rev. Lett. {\bf 89} (2002) 075505. 
\bibitem{Peng} L.-M. Peng, Z. L. Zhang, Z. Q. Xue, Q. D. Wu, 
Z. N. Gu and D. G. Pettifor, Phys. Rev. Lett. {\bf 85} (2000) 3249. 
\bibitem{Zhao}Y. Zhao, R. E. Smally and B. I. Yakobson, 
Phys. Rev. B {\bf 66} (2002) 195409. 
\bibitem{Struik} D. J. Struik, {\it Lectures on 
Classical Differential Geometry} 
(Addison-Wesley, 1950, Massachusetts). 
\end{thebibliography}
\end{document}